\documentclass{article}

\usepackage{graphicx}
\usepackage{amssymb}
\usepackage{tabularx}
\usepackage[utf8]{inputenc}
\usepackage[english]{babel}
 \usepackage{amsthm}
\usepackage{amsmath,amsfonts}
\usepackage[usenames, dvipsnames]{color}
\usepackage{dsfont}
\usepackage{bbm}
\usepackage{breqn}
\usepackage{amssymb}
\usepackage{url}
\usepackage[utf8]{inputenc}
\usepackage{float}

\usepackage{amssymb}
\usepackage{tabularx}
\usepackage[utf8]{inputenc}
\usepackage[english]{babel}
 \usepackage{amsthm}
\usepackage{amsmath,amsfonts}
\usepackage[usenames, dvipsnames]{color}
\usepackage{xcolor}
\usepackage{dsfont}
\usepackage{bbm}
\usepackage{breqn}
\usepackage{amssymb}
 \usepackage{lscape}
\usepackage{bbold}
\usepackage{url}
\usepackage{bm}

\usepackage{amssymb}
\usepackage{tabularx}
\usepackage[utf8]{inputenc}
\usepackage[english]{babel}
 \usepackage{amsthm}
\usepackage{amsmath,amsfonts}
\usepackage[usenames, dvipsnames]{color}
\usepackage{dsfont}
\usepackage{bbm}
\usepackage{breqn}
\usepackage{amssymb}
 \usepackage{lscape}
\usepackage{bbold}
\usepackage{url}
\usepackage{ulem}

\usepackage{array,ragged2e}

\theoremstyle{definition}
\newtheorem{definition}{Definition}

\usepackage{xcolor}

\title{A spatio-temporal hybrid Strauss hardcore point process for forest fire occurrences}

\author{Morteza Raeisi, Florent Bonneu, Edith Gabriel \\ 
LMA (Avignon University) \& BioSP (INRAE)
}

\begin{document}

\maketitle


\begin{abstract}
We propose a new point process model that combines, in the spatio-temporal setting, both multi-scaling by hybridization and hardcore distances. Our so-called hybrid Strauss hardcore point process model allows different types of interaction, at different spatial and/or temporal scales, that might be of interest in  environmental and biological applications. The inference and simulation of the model are implemented using the logistic likelihood approach and the  birth-death Metropolis-Hastings algorithm. Our model is illustrated and compared to others on two datasets of forest fire occurrences respectively in France and Spain.
\end{abstract}

\section{Introduction}
\label{intro}
In point process modeling, most of the existing models yield point patterns with mainly single-structure, but only a few with multi-structure. Interactions with single-structure are often classified into three classes: randomness, clustering and inhibition. 
Among the inhibition processes is the  hardcore process. It has some hardcore distance $h$ in which distinct points are not allowed to come closer than a distance $h$ apart. 
This type of interaction  can be  modeled by Gibbs point processes as the hardcore or Strauss hardcore point processes and also by Cox point processes as Matérn’s hardcore (Matérn, 1960; 1986) or Matérn thinned Cox point processes (Andersen and Hahn, 2016). Here, we focus on the former, i.e. Gibbs models implemented by a hardcore component as in the Strauss hardcore model. The  form of Strauss hardcore density indicates that the hardcore parameter only rules at least the distance between points, and has no effect on the interaction terms of the density (Dereudre and  Lavancier, 2017, sect. 2.3).  

In several domains, there exist point patterns with hardcore distances that have to be modeled.
Spatial point  patterns with hardcore property  can be found in capillaries studies (Mattfeldt et al., 2006; 2007; 2009), in texture synthesis (Hurtut et al., 2009), in forest fires (Turner, 2009), in cellular networks (Taylor et al., 2012 and Ying et al., 2014), in landslides (Das and Stein, 2016), in modern and contemporary architecture and art (Stoyan, 2016) and in location clustering econometrics (Sweeney and Gomez-Antonio, 2016). 

There also exist  point  patterns with either clustering and inhibition like hardcore 
interactions at different scales simultaneously (Badreldin et al., 2015; Andersen and Hahn, 2016 and Wang et al., 2020). Wang et al. (2020, sect. 2.4) investigated the effect of the hardcore distance  on spatial patterns of trees by comparing the pair correlation function curves for different values of hardcore distances in the  fitted hybrid Geyer hardcore model. 
Raeisi et al. (2019) review spatial and spatio-temporal point processes that model both inhibition and clustering at different scales. Such multi-structure interactions can be modeled by the spatial hybrid Gibbs point process (Baddeley et al, 2013).  In this paper,  we aim to extend the spatial Strauss hardcore point process (Ripley, 1988) to the spatio-temporal framework and introduce a multi-scale version of it using a hybridization approach. We use this model to
describe one of the most complex phenomena from the spatio-temporal modeling point of view: forest fire occurrences.

 The complexity of forest fire occurrences is due in particular to the existence of multi-scale structures and hardcore distances in space and time.
For instance, spatio-temporal variations of fire occurrences depend on the spatial distribution of current land use and weather conditions.
Changes in vegetation due to forest fires burnt areas further affect the probability of fire occurrences during the regeneration period leading to the existence of hardcore distances in space-time. The multi-scale structure of clustering and inhibition in the spatio-temporal pattern of forest fire occurrences is discussed in  Gabriel et al. (2017).
Wildfires have mainly been modeled by Cox processes and inferred by Bayesian hierarchical approaches, as the integrated nested Laplace approximation (INLA) approach (Rue et al., 2009). See Møller and Diaz-Avalos (2010),   Pereira et al. (2013), Serra et al. (2012, 2014a,b),  Najafabadi et al. (2015), Juan (2020) and Pimont et al. (2021) for single-structure models and  Gabriel et al. (2017), Opitz et al. (2020) for multi-structure models.   Recently, Raeisi et al. (2021) modeled the multi-structure of forest fire occurrences by a spatio-temporal Gibbs process and use a composite likelihood approach for its inference.

This paper is organized as follows. 
In Section $2$ we introduce in the spatio-temporal framework the notations and definitions of Gibbs point processes in order to introduce our multi-scale version of the Strauss hardcore model. Section $3$ is devoted to the inference of our model. It describes techniques to determine the irregular parameters (hardcore and interaction distances) and the logistic-likelihood approach generalized to the spatio-temporal setting to estimate the regular parameters (strength of interactions). Section $4$ illustrates the goodness-of-fit of the logistic likelihood approach on simulated patterns of our model obtained by an extended Metropolis-Hastings algorithm.
 Finally, in Section~\ref{sec:application}, we illustrate our model on two monthly and yearly records of forest fires in France and Spain and compare our results to those obtained by two other spatial point process models.

\section{Towards multi-scale Strauss hardcore point processes}

Gibbs models are flexible point processes that allow the specification of point interactions via a probability density  defined with respect to the unit rate Poisson point process. These models allow us to characterize a form of local or
Markovian dependence amongst events. Gibbs point processes contain a large class of  flexible and natural models that can be applied for:
\begin{itemize}
 \item   Postulating  the interaction mechanisms between pairs of points, 
\item   Taking into account  clustering, randomness, or  inhibition structures,
 \item  Combining several structures at different scales with the hybridization approach.
\end{itemize}

Let $\textbf{x}=\{(\xi_1,t_1),,...,(\xi_n,t_n)\}$ be a spatio-temporal point pattern where $(\xi_i,t_i)\in W=S \times T \subset \mathds{R}^2\times \mathds{R}$.  We consider $(W,d(\cdot,\cdot))$ where $d((u,v),(u',v')):=\max\{||u-u'||,|v-v'|  \}$ for $(u,v),(u',v') \in    W$ is a complete, separable metric space.  The cylindrical neighbourhood $C_r^q(u,v)$ centred at $(u,v)\in W$ is defined by
\begin{equation}
C_r^q(u,v)=\{ (a,b)\in W : ||u-a||\leq r ,|v-b|\leq q \},
\end{equation}
where $r,q>0$ are spatial and temporal radius and  $||\cdot||$ denotes the Euclidean distance in $\mathds{R}^2$ and $|\cdot|$ denotes the usual distance in $\mathds{R}$. Note that $C_r^q(u,v)$ is a cylinder
with centre $(u,v)$, radius $r$, and height $2q$.

 A finite Gibbs point process is a
finite simple point process defined with a density $f(\textbf{x})$ on $W$ that satisfies the hereditary condition, i.e. $f(\textbf{x})>0 \Rightarrow f(\textbf{y})>0$ for all $\textbf{y} \subset \textbf{x}$. Throughout the paper, we suppose that $W$ is a bounded set in $\mathds{R}^2\times \mathds{R}$, and all Gibbs models are defined with respect to a unit rate Poisson point process on $W$.

A closely related concept to density functions is Papangelou conditional intensity function (Papangelou, 1974) which is  a characterization of Gibbs point processes useful for inferring the model parameters and simulating related point patterns. The Papangelou conditional intensity of a spatio-temporal point process on $W$ with density $f$  is defined, for $(u,v)\in  W $, by 
\begin{equation}
\lambda((u,v)|\textbf{x}) = \frac {f(\textbf{x} \bigcup  (u,v))}{f(\textbf{x}\backslash (u,v))},
\end {equation}
with $a/0 := 0$ for all $a\geq 0$ (Cronie and van Lieshout, 2015). 

 The Papangelou conditional intensity is very useful to describe local interactions in a point pattern and leads to the notion of a Markov point process which is the basis for the implementation of MCMC algorithms used for simulating of Gibbs models. We say that
the point process has "interactions of range $R$ at $(\xi,t)$" if points further
than $R$ away from $(\xi,t)$ do not contribute to the conditional intensity at $(\xi,t)$.  A spatio-temporal Gibbs point process $X$  has a finite interaction range $R$ if the Papangelou
conditional intensity satisfies
\begin{equation}
\lambda((u,v) | \textbf x) = \lambda((u,v)| \textbf x \cap C_R^R(u,v)) 
\label{eq.3258}
\end{equation}
for all configurations $\textbf x$ of $X$ and all $(u,v)\in W,$ where $C_R^R(u,v)$ denotes the cylinder of radius $R > 0$ and height $2R>0$ centered at $(u,v)$ (see Iftimi et al. (2018) for a well-detailed discussion). Spatio-temporal Gibbs models usually have finite interaction range property (spatio-temporal Markov property) and are called in this case Markov point processes (van Lieshout 2000). The finite range property of a spatio-temporal Gibbs model implies that the probability to insert a point $(u,v)$ into $\textbf x$ depends only on some cylindrical
neighborhood of $(u,v)$.   Finally, a spatio-temporal Gibbs point process is said to be locally stable if  is stochastically dominated by a Poisson point processs, that is if there exists $\lambda_{constant}<\infty$ such that for any $(u,v)\in W$ and for any configuration $\mathbf{x}$ of $X$, $\lambda((u,v)|\mathbf{x})\leq \lambda_{constant}$. We further refer to Dereudre (2019)  for a more formal introduction of Gibbs point processes.

Here, we first review  spatio-temporal Gibbs models and then extend the spatial Strauss hardcore model to the spatio-temporal and multi-scale context.

\subsection{Single-scale Gibbs point process models}

In the literature, several spatio-temporal Gibbs point process models have been
proposed such as the hardcore (Cronie and van Lieshout, 2015), Strauss (Gonzalez
et al., 2016), area-interaction (Iftimi et al., 2018), and Geyer (Raeisi et al., 2021) point processes. 

A Gibbs point process model explicitly postulates that interactions traduce dependencies between the points of the pattern. The hardcore interaction is one of the simplest type of interaction, which forbids points being too close to each other. The homogeneous spatio-temporal hardcore point process is defined by the density
\begin{equation}
f(\textbf{x})= c\lambda^{n(\textbf{x})}\mathbb{1}\{||\xi-\xi'||>h_s \text{ or }|t-t'|>h_t;\forall (\xi,t)\neq (\xi',t')\in \textbf{x}\},
\label{eq.22058}
\end{equation}
 where $c > 0$ is a normalizing constant, $\lambda  > 0$ is an activity parameter, $h_s,h_t> 0$ are, respectively, the spatial and the temporal hardcore distances and $n(\textbf{x})$ is the number of points in $\textbf{x}$. The 
Papangelou conditional intensity  of a homogeneous spatio-temporal hardcore point process
 for $(u,v) \notin \textbf{x}$ is obtained 
\begin{align}
\begin{split}
\lambda((u,v)|\textbf{x})&= \lambda\mathbb{1}\{||\xi-u||>h_s  \text{ or } |t-v|>h_t;\forall (\xi,t)\in \textbf{x}\}\\
&=\lambda \prod_{(\xi,t)\in \textbf{x}}\mathbb{1}\{||\xi-u||>h_s\text{ or }|t-v|>h_t\}\\
&=\lambda \prod_{(\xi,t)\in \textbf{x}}\mathbb{1}\{(\xi,t)\notin C_{h_s}^{h_t}(u,v)\}.
\end{split}
\end{align}
The homogeneous  spatio-temporal Strauss point process is defined by density
\begin{equation}
f(\textbf{x})= c\lambda^{n(\textbf{x})}\gamma^{S_r^q(\textbf{x})},
\label{eq.0258}
\end{equation}
 where $0<\gamma\leq1$, $$S_r^q(\textbf{x})=\sum_{(\xi,t)\neq(\xi',t')\in\textbf{x}}\mathbb{1}\{||\xi-\xi'||\leq r,|t-t'|\leq q\}$$ and the Papangelou conditional intensity of the model is
\begin{equation}
 \text{for } (u,v) \notin \textbf{x}, \ \ \lambda((u,v)|\textbf{x})=\lambda \gamma^{n[C_r^q(u,v);\textbf{x}]},
\end{equation}
and 
\begin{equation}
\text{for } (\xi,t) \in \textbf{x}, \ \ \lambda((\xi,t)|\textbf{x})=\lambda \gamma^{n[C_r^q(\xi,t);\textbf{x} \setminus (\xi,t) ]},
\end{equation}
where $n[C_{r}^{q}(y,z);\textbf{x}]=\sum _ {(\xi,t)\in \textbf{x} }  \mathbb{1} \{||y-\xi||\leq r, |z-t|\leq q\}$ is the number of points in $\textbf{x}$  which are in a cylinder $C_r^q(y,z)$.
Although the Strauss point process was originally intended as a model of
clustering, it can only be used to model inhibition, because the parameter $\gamma$
cannot be greater than 1. If we take 
 $\gamma> 1$, the density function of Strauss model is not integrable,
so it does not define a valid probability density.

\medskip

As mentioned, the Strauss point process  model only achieves the inhibition structure. In the spatial framework, two ways are introduced to overcome this problem that we extend to the spatio-temporal framework hence defining two new spatio-temporal Gibbs point process models. 

A first way is to consider an upper bound  for the number of
neighboring points that interact. In this case, Raeisi et al. (2021) defined a homogeneous spatio-temporal  Geyer saturation  point process by density
\begin{equation}
f(\textbf{x})=c \lambda^{n(\textbf{x})} \prod_{(\xi,t)\in \textbf{x}} \gamma^ {min\{s,n^*[C_{r}^{q}(\xi,t);\textbf{x}]\}},
\end{equation}
 where $s$ is a saturation parameter and 
$n^*[C_{r}^{q}(\xi,t);\textbf{x}]=n[C_{r}^{q}(\xi,t);\textbf{x}\setminus (\xi,t)]=\sum _ {(u,v)\in \textbf{x} \backslash (\xi,t)}  \mathbb{1} \{||u-\xi||\leq r, |v-t|\leq q\}.$

A second way is to introduce a hardcore condition to the Strauss density (\ref{eq.0258}). Hence, we can define a Strauss hardcore model in the spatio-temporal context.
\theoremstyle{definition}
\begin{definition}
We define the  \textit{spatio-temporal Strauss hardcore point process} as the point process  with density
\begin{align}
\begin{split}
f(\textbf{x})&= c\lambda^{n(\textbf{x})}\gamma^{S_r^q(\textbf{x})}\mathbb{1}\{||\xi-\xi'||>h_s \text{ or }|t-t'|>h_t;\forall (\xi,t)\neq (\xi',t')\in \textbf{x}\},
\end{split}
\label{Eq.01236}
\end{align}
where $0<h_s<r$, $0<h_t<q$ and $\gamma>0$. 
\end{definition}
Note that, contrary to the case of a spatio-temporal 
Strauss process (\ref{eq.0258}), for Strauss hardcore processes the interaction
parameter $\gamma$ can assume any non-negative value, in particular
a value larger than 1. If the interaction parameter $\gamma$ is bigger than one, we have
an attraction effect, whereas for $\gamma$ smaller than one there is a repulsion
tendency. If $\gamma=1$ a spatio-temporal hardcore process (\ref{eq.22058}) is obtained. 

 The Papangelou conditional intensity of a homogeneous spatio-temporal Strauss hardcore point process for $(u,v) \notin \textbf{x}$ is obtained
\begin{align}
\begin{split}
\lambda((u,v)|\textbf{x})&= \lambda\gamma^{n[C_r^q(u,v);\textbf{x}]}\mathbb{1}\{||\xi-u||>h_s\text{ or }|t-v|>h_t;\forall (\xi,t)\in \textbf{x}\}\\
&=\lambda \gamma^{n[C_r^q(u,v);\textbf{x}]} \prod_{(\xi,t)\in \textbf{x}}\mathbb{1}\{(\xi,t)\notin C_{h_s}^{h_t}(u,v)\}.
\end{split}
\end{align}

We can define inhomogeneous versions of all above models by replacing the constant $\lambda$ by a function $\lambda(\xi,t)$, inside the product operator over $(\xi,t) \in \textbf{x}$, that expresses a spatio-temporal trend, which can be a function of the coordinates of the points and depends on covariate information.

\subsection{Multi-scale Gibbs point process models}

Since most natural phenomena exhibit dependence at multiple scales as earthquake (Siino et al., 2017;2018) and forest fire occurrences (Gabriel et al., 2017), single-scale Gibbs point process models are unrealistic in many applications. This motivates us and other statisticians to construct multi-scale
generalizations of the classical Gibbs models.  Baddeley et al. (2013) proposed hybrid models as a general way to generate multi-scale processes combining Gibbs processes.
 Given $m$  densities $f_1,f_2, . . .,f_m$ of Gibbs point processes, the hybrid
density is defined as $f(\textbf{x})=cf_1(\textbf{x})\times f_2(\textbf{x})\times \cdots \times f_m(\textbf{x})$ where $c$ is a normalization constant.  

Iftimi et al. (2018) extended the hybrid approach for an area-interaction model to the spatio-temporal framework where the density is given by
\begin{equation}
f(\textbf{x})=c \prod_{(\xi,t)\in \textbf{x}} \lambda(\xi,t)  \prod_{j=1}^{m}\gamma_j^ {-\ell(\cup_{(\xi,t)\in\textbf{x}} C_{r_j}^{q_j}(\xi,t))},
\end{equation}
 where $(r_j, q_j)$ are pairs of irregular parameters  of the model and $\gamma_j$ are interaction parameters, $j = 1, \dots ,m$.

In the same way, Raeisi et al. (2021) defined a  spatio-temporal multi-scale Geyer saturation point process with density
\begin{equation}
f(\textbf{x})=c \prod_{(\xi,t)\in \textbf{x}} \lambda(\xi,t) \prod_{j=1}^{m}\gamma_j ^ {min\{s_j,n^*[C_{r_j}^{q_j}(\xi,t);\textbf{x}]\}}
\label{Eq.7}
\end{equation}
where $c > 0$ is a normalizing constant, $\lambda \geq 0$ is a measurable and bounded function, $\gamma_j> 0$ are the interaction parameters.

Similarly, a hybrid version of spatio-temporal Strauss model can be defined by hybridization.
\theoremstyle{definition}
\begin{definition}
We  define the \textit{spatio-temporal hybrid Strauss point process} with density
\begin{equation}
f(\textbf{x})= c\prod_{(\xi,t)\in \textbf{x}} \lambda(\xi,t) \prod_{j=1}^{m}\gamma_j^{S_{r_j}^{q_j}(\textbf{x})},
\label{Eq.1127845895}
\end{equation}
 where $0<\gamma_j<1$ for all $j=1,...,m$.
\end{definition}
Note that we called the model (\ref{Eq.1127845895}) hybrid rather than multi-scale. The model  (\ref{Eq.1127845895}) can cover inhibition structure because $0< \gamma_j < 1, \forall j \in \{1,\dots,m\}$. However, it can take into account clustering if one of densities in hybrid is the one of a hardcore process. 

\subsection{Hybrid Strauss hardcore point process }

The hybrid Gibbs point process models do not necessarily include $m$ same Gibbs point process models (see Baddeley et al., 2015 sect. 13.8). Badreldin et al. (2015) applied a spatial hybrid model including a hardcore density to model strong inhibition at very short distances, Geyer density for cluster structure in short to medium distances and a Strauss density for a randomness structure in larger distances to the spatial pattern of the halophytic species distribution in an arid coastal environment. Wang et al. (2020) fitted a spatial hybrid Geyer hardcore point process on the tree spatial distribution patterns.  In this section, we extend this type of hybrids to the spatio-temporal context. 

\theoremstyle{definition}
\begin{definition}
We define the  \textit{spatio-temporal hybrid Strauss hardcore point process}  with density
\begin{align}
\begin{split}
f(\textbf{x})&= c\prod_{(\xi,t)\in \textbf{x}} \lambda(\xi,t) \prod_{j=1}^{m}\gamma_j^{S_{r_j}^{q_j}(\textbf{x})}\\
&\hspace{.3cm}\times\mathbb{1}\{||\xi'-\xi''||>h_s \text{ or }|t'-t''|>h_t;\forall (\xi',t')\neq (\xi'',t'')\in \textbf{x}\},
\end{split}
\label{Eq.154215}
\end{align}
where $0<h_s<r_1< \cdots < r_m$, $0<h_t<q_1< \cdots < q_m$ and $\gamma_j>0$ for all $j=1,...,m$.
\end{definition}
The model could be used to model clustering patterns with a softer attraction between the points like a pattern with a combination of interaction terms that show repulsion between the points at a small scale and attraction between the points at a larger scale.

The Papangelou conditional intensity of an inhomogeneous spatio-temporal hybrid Strauss hardcore process is then,
for $(u,v) \notin \textbf{x}$, 
\begin{align}
\begin{split}
\lambda((u,v)|\textbf{x})&= \lambda(u,v)\prod_{j=1}^{m} \gamma_j^{n[C_{r_j}^{q_j}(u,v);\textbf{x}]}\mathbb{1}\{||\xi-u||>h_s \text { or }|t-v|>h_t;\forall (\xi,t)\in \textbf{x}\}\\
&=\lambda(u,v)\prod_{j=1}^{m} \gamma_j^{n[C_{r_j}^{q_j}(u,v);\textbf{x}]}\prod_{(\xi,t)\in \textbf{x}}\mathbb{1}\{(\xi,t)\notin C_{h_s}^{h_t}(u,v)\}.
\end{split}
\label{Eq.1023560}
\end{align}
Because, the conditional intensity of Gibbs models including a hardcore interaction term takes the value zero at some locations, we can rewrite it as
\begin{equation}
\lambda((u,v)|\textbf x)=m((u,v)|\textbf x)\lambda^+((u,v)|\textbf x),
\label{Eq.201322}
\end{equation}
where $m((u,v)|\textbf x)$ takes only the values 0 and 1, and $\lambda^+((u,v)|\textbf x)>0$   everywhere.

The spatio-temporal hybrid Strauss hardcore point process (\ref{Eq.154215}) 
is a Markov point process in Ripley-Kelly’s (1977) sense at interaction range $ \max\{r_m, q_m\}$ and is also locally stable. These can be shown as in Iftimi et al. (2018) and Raeisi et al. (2021).  The Markov property allows us to design MCMC simulation algorithms and local stability property ensures the convergence of simulation algorithms. The spatio-temporal hybrid Strauss hardcore point process (\ref{Eq.154215}) is well-defined due to the log-linearity of its  conditional intensity (see Section \ref{sec:inference})   and satisfying in Markov  and local stability properties (Vasseur et al., 2020) and moreover, 
it exists also in $\mathds{R}^d\times \mathds{R}$ (Dereudre 2019).

\section{Inference}
\label{sec:inference}

Gibbs point process models involve two types of parameters: regular 
and irregular parameters. A parameter is called regular if the log likelihood of density is a linear function of that parameter otherwise it is called irregular.
 Typically, regular parameters determine the ‘strength’ of the interaction, while irregular parameters
determine the ‘range’ of the interaction. As an example, in the Strauss hardcore point process (\ref{Eq.01236}), the trend parameter  $\lambda$ and the interaction $\gamma$ 
 are regular parameters and   the interaction distances $r$ and $q$ and the hardcore distances $h_s$ and $h_t$ are irregular parameters.

To determine the 
interaction distances $r$ and $q$, there are several practical techniques, but no general statistical theory available. A useful technique is the maximum profile pseudo-likelihood approach (Baddeley and Turner, 2000). In the spatio-temporal framework, Iftimi et al. (2018) and Raeisi et al. (2021) 
 selected feasible range of irregular parameters by analyzing the behavior of some summary statistics and the goodness-of-fit of several models with different combinations of parameters.
 
\begin{figure}
\begin{center}
\includegraphics[width=5cm,height=4cm]{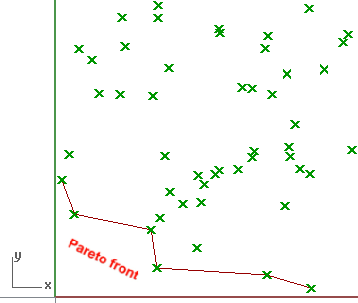}
\caption{The red line is an example of a Pareto front where the frontier and the area below it are a continuous set of choices. The points on the red line are examples of Pareto-optimal choices of hardcore parameter estimates.}
\label{c4:fig1}
\end{center}
\end{figure}

The hardcore interaction term $m(\cdot|\textbf x)$ in the conditional intensity (\ref{Eq.201322}) does not depend on the other parameters of the densities of Gibbs point processes. This implies that it can first be estimated and kept fixed for the sequel (Baddeley et al., 2019, p. 26). In the spatial framework, the maximum likelihood estimate of the hardcore distance in $m(\cdot|\textbf x)$ corresponds to the minimum interpoint distance (Baddeley et al., 2013, Lemma 7). The generalization to the spatio-temporal context with a cylindrical hardcore structure implies to consider a multi-objective minimization problem over the spatial and temporal hardcore distances $h_s$ and $h_t$. The choice of our hardcore parameters needs to analyze the Pareto front of feasible solutions on the graph of spatial and temporal interpoint distances. We refer the reader to Ehrgott (2005) for a description of multi-criteria optimization and the definition of Pareto optimality. To estimate the hardcore distance $h_s$ and $h_t$, we consider a feasible solution on the Pareto front as large as possible and with a ratio consistent with our knowledge of interaction mechanisms in practice. Indeed, in the spatial
case, there is a unique solution for the minimum interpoint distance. In the spatio-temporal
case, we search $h_s$ and $h_t$ minimizing respectively the interpoint distances
in space and time for a same couple of points. Because the couple $(h_s; h_t)$ does not
exist in general. we thus have to consider the feasible solutions on the Pareto front (see Figure \ref{c4:fig1}).

 Regular parameters can be estimated using the pseudo-likelihood method (Baddeley and Turner, 2000) or logistic likelihood method (Baddeley et al., 2014) rather than the maximum likelihood method (Ogata and Tanemura, 1981). Due to the advantage of the logistic likelihood over pseudo-likelihood for spatio-temporal Gibbs point processes (Iftimi
et al., 2018; Raeisi et al., 2021), we implement the former approach in Raeisi et al. (2021,  \textit{Algorithm 2}) for regular parameter estimation   of the spatio-temporal hybrid Strauss hardcore point process.

We assume that $\boldsymbol\theta = (\log \gamma_1, \log \gamma_2, \dots, \log \gamma_m)$ is
 the logarithm of interaction parameters in spatio-temporal hybrid Strauss hardcore point process (\ref{Eq.154215}). 
To estimate  $ \boldsymbol\theta$, due to (\ref{Eq.201322}), we just consider the points $(u,v)$ where $m((u,v) | \textbf{x})$ is equal to $1$ in (\ref{Eq.1023560}). 
By defining $S_j((u,v),\textbf{x}):=n[C_{r_j}^{q_j}(u,v);\textbf{x}\setminus (u,v)]$ in (\ref{Eq.1023560}), we can thus write $\lambda_{\boldsymbol\theta}((u,v)|\textbf{x})= \lambda (u,v) \prod_{j=1}^{m} \exp(\theta_j S_j ((u,v),\textbf{x}))$.
Hence,  the logarithm of the Papangelou conditional intensity of the  spatio-temporal  hybrid Strauss hardcore  point process 
for $(u,v) \in W $ which satisfies in hardcore condition, i.e. $m((u,v) | \textbf{x})=1$ in (\ref{Eq.1023560}), is
\begin{align}
\begin{split}
\log \lambda((u,v)|\textbf{x}) &= \log  \lambda (u,v) + \sum_{j=1}^{m} (\log \gamma_j ) S_j ((u,v),\textbf{x})\\
&= \log  \lambda (u,v) +\boldsymbol \theta^{\top} \boldsymbol S((u,v),\textbf{x})
\end{split}
\end{align}
corresponding to a linear model in $\boldsymbol \theta$ with offset $\log  \lambda (u,v)$ where $\boldsymbol S((u,v),\textbf{x})=[S_1((u,v),\textbf{x}), S_2((u,v),\textbf{x}),...,S_m((u,v),\textbf{x})]^{\top}$ is a sufficient statistics.

By considering a set of dummy points $\textbf{d}$ from an independent Poisson process with intensity function $\rho$, we obtain by defining the Bernoulli variables  $Y((\xi,t))=\mathbbm{1}_{\{(\xi,t)\in \textbf{x}\}}$ for $(\xi,t)\in \textbf{x}\cup \boldsymbol d$ that the logit of $P(Y((\xi,t)))=1$ is equal to $\log \frac {\lambda_{\boldsymbol \theta}((\xi,t)|\textbf x \backslash (\xi,t))}{\rho (\xi,t)}$. Under regularity conditions, conditional on $\textbf{x}\cup \textbf{d}$, the log-logistic likelihood
\begin{align}
\begin{split}
\log LL(\textbf {x},\textbf d;\boldsymbol \theta)&=\sum_{(\xi,t)\in \textbf {x}} \log \frac {\lambda_{\boldsymbol \theta}((\xi,t)|\textbf {x} )}{\lambda_{\boldsymbol \theta}((\xi,t)|\textbf {x} )+\rho (\xi,t)}
\\
&+\sum_{(\xi,t)\in \textbf d} \log \frac {\rho(\xi,t)}{\lambda_{\boldsymbol \theta}((\xi,t)|\textbf {x} )+\rho (\xi,t)},
\label{Eq.2511112}
\end{split}
\end{align}
 corresponds to the
logistic regression model with responses $Y((\xi,t))=\mathbbm{1}_{\{(\xi,t)\in \textbf{x}\}}$ and offset term
$-\log \rho(\xi,t)$ and admits a unique maximum which already implemented by using standard software for GLMs.

The term $\lambda(\xi,t)$ may also depend on a
parameter, say $\boldsymbol {\beta}$. We shall consider $\lambda(\xi,t)=\beta$ in the simulation study and $\lambda(\xi,t)=\exp (\boldsymbol {\beta}^{\intercal} \boldsymbol Z(\xi,t))$ where $\boldsymbol Z(\xi,t)$ is a matrix of covariates. 
In summary we have the following algorithm for data and dummy points such that $m(\cdot|\textbf x)=1$.

\vspace{.2cm}

\textbf{\textit{Algorithm}}
\vspace{.1cm}
\begin{itemize}

\item Generate a set of dummy points according to a Poisson process with intensity
function $\rho$ and merge them with all the data points in $\textbf x$ to construct the set of quadrature
points $(\xi_k, t_k) \in  W=S \times T$,

\item Obtain the response variables $y_k$ (1 for data points, 0 for dummy points),

\item  Compute the values $\boldsymbol S((\xi_k, t_k),\textbf{x})$ of the vector of sufficient statistics at each quadrature point,

\item Fit a logistic regression model with explanatory variables $\boldsymbol S((\xi_k, t_k),\textbf{x})$ and $ \boldsymbol Z(\xi_k,t_k)$,
and offset $\log (1/\rho(\xi_k, t_k))$ on the  responses $y_k$ to obtain estimates $\hat{\boldsymbol \theta}$ for the $\boldsymbol S$-vector and  $\hat{\boldsymbol {\beta}}$ for covariate-vector,

\item Return the parameter estimator $\hat{\boldsymbol \gamma}=\exp (\hat{\boldsymbol \theta})$ and $\hat{\boldsymbol{\beta}}$.

\end{itemize}

\section{Simulation study}

Due to the markovian property of the spatio-temporal hybrid Strauss hardcore point process (\ref{Eq.154215}), its Papangelou conditional intensity at a point thus depends only on that point and its neighbors in $\textbf{x}$. Hence, we can design simulation approach by Markov chain Monte Carlo algorithms.

 Gibbs point process models can be
simulated a birth-death Metropolis-Hastings algorithm that typically requires only computation
of  the Papangelou conditional intensity (Møller and Waagepetersen, 2004). 
Raeisi et al. (2021) extended the  birth-death
Metropolis-Hastings algorithm to the spatio-temporal context that we adapt here for simulating the spatio-temporal hybrid Strauss hardcore point process. 

We implement the estimation and simulation algorithms in \texttt{R} (R Core Team, 2016) and generate  simulations of three stationary spatio-temporal hybrid Strauss hardcore point processes specified
by a conditional intensity of the form (\ref{Eq.1023560}) 
in $W =[0, 1]^3$. The parameter values used for the simulations are reported in Table \ref{Tab1}. 
\begin{table}
\centering
\caption{Parameter combinations of three hybrid Strauss hardcore point process models used in simulation study.}
 \label{Tab1}
\small
\begin{tabular}{cccccc}
\hline
& \multicolumn{5}{c}{Values of parameter}\\
\cline{2-6}
          &   \multicolumn{2}{c}{Regular parameters}  & & \multicolumn{2}{c}{Irregular parameters}  \\
\cline{2-3}\cline{5-6}
Model & $\lambda$ & $\gamma$  & & $r,q$ & $h_s,h_t$ \\
\hline
\textit{Model 1}   & 70 & (0.8,0.8)   & &  (0.05,0.1)  &  (0.01,0.01) \\
\textit{Model 2}   & 50 &  (1.5,1.5)   & &   (0.05,0.1) & (0.01,0.01) \\
\textit{Model 3}   & 70 & (0.5,1.5)  & &   (0.05,0.1)  & (0.01,0.01)\\
\hline
\end{tabular}
\end{table}
The spatial and temporal radii $r$ and $q$, spatial and temporal hardcores $h_s$ and $h_t$, 
are treated as  known parameters. 

\begin{figure}
\begin{center}
\includegraphics[width=7cm,height=3.5cm]{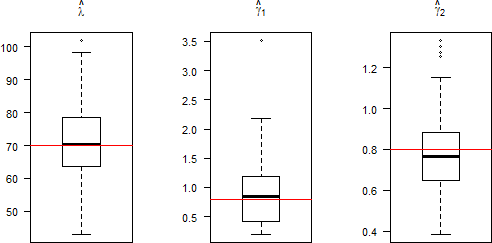}
\includegraphics[width=7cm,height=3.5cm]{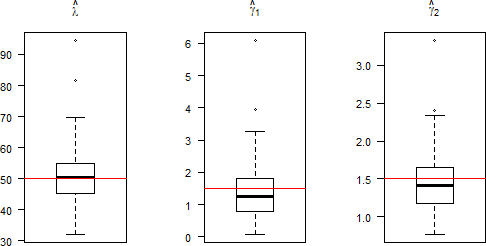}
\includegraphics[width=7cm,height=3.5cm]{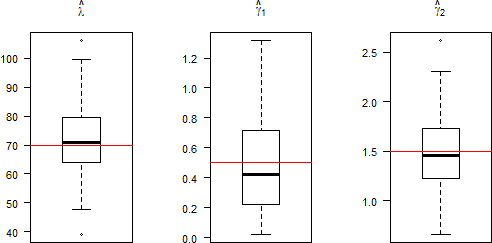}
\caption{\footnotesize{Boxplots of regular parameter estimates of the hybrid Strauss hardcore point process  obtained from the logistic likelihood estimation methods. Up to down: \textit{Model 1},  \textit{Model 2}, and \textit{Model 3} }}
\label{pic1}
\end{center}
\end{figure}

We generate 100 simulations of each specified model.
Boxplots of parameter estimates $\lambda$, $\gamma_1$, and $\gamma_2$  obtained from the logistic likelihood estimation method for each model are shown in Figure \ref{pic1}. 
The red horizontal lines represent the true parameter values. Point and interval  parameter estimates $\lambda$, $\gamma_1$, and $\gamma_2$ are reported in Table \ref{Tab3}. 
Most of 
the estimated parameter values are close to the true values for three models. Due to visual and computational comparisons, we conclude that the logistic likelihood approach performs well for spatio-temporal hybrid Strauss hardcore point processes. 

\begin{table}
\centering
\caption{Mean and $95\%$ interval regular parameter estimates of the three hybrid Strauss hardcore point process models used in simulation study.}
 \label{Tab3}
\small
\begin{tabular}{ccc}
\hline
True values & Mean & $95\%$ CI   \\
\hline
          &   \multicolumn{2}{c}{\textit{Model 1}}  \\
\cline{2-3}
$\lambda=70$   &71.43&(69.16,73.70) \\
$\gamma_1=0.8$    &0.89&(0.78,1.00) \\
$\gamma_2=0.8$    &0.78&(0.74,0.82) \\
\hline
       &      \multicolumn{2}{c}{\textit{Model 2}}  \\
          \cline{2-3}
          $\lambda=50$ &50.84&(48.99,52.68) \\
          $\gamma_1=1.5$ &1.41&(1.23,1.60) \\
           $\gamma_2=1.5$&1.46&(1.38,1.54)\\
          \hline
             &   \multicolumn{2}{c}{\textit{Model 3}}  \\     
             \cline{2-3}
  $\lambda=70$ &71.67&(69.18,74.15)  \\ 
  $\gamma_1=0.5$&0.50 &(0.43,0.57)\\
   $\gamma_2=1.5$&1.49&(1.42,1.55)\\
  \hline
\end{tabular}
\end{table}

\section{Application}
\label{sec:application}

In this section, we aim to model the interactions of forest fire occurrences across a range of spatio-temporal scales. We compare the relevance of our model on two datasets in the center of Spain and south of France, w.r.t. the ones of two widely used models: the inhomogeneous Poisson process and the log-Gaussian Cox process (LGCP).

\subsection{Data description} 

\textbf{\texttt{clmfires} dataset}

\begin{figure}
\begin{center}
\includegraphics[width=12cm,height=3cm]{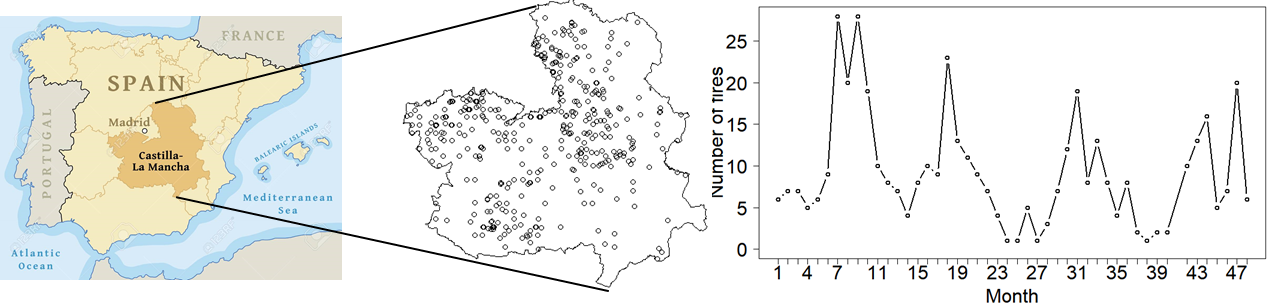}
\caption{\footnotesize{Left: Map of  the region of Castilla-La Mancha (Spain). Middle: Forest fire locations. Right: monthly number of  fires recorded between January $2004$ and December $2007$ with burnt areas, spatial distances, and time distances respectively bigger than 5 $ha$, 0.2 $km$, and 100 $days$.}}
\label{fig:clm}
\end{center}
\end{figure}

\begin{figure}
\begin{center}
\includegraphics[width=3.25cm,height=3cm]{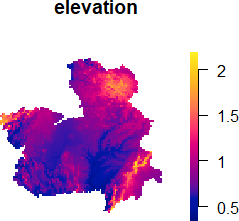}
\includegraphics[width=3.25cm,height=3cm]{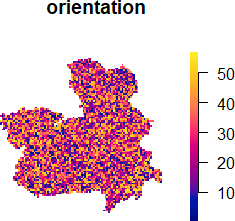}
\includegraphics[width=3.25cm,height=3cm]{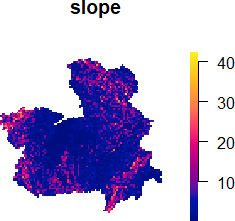}

\includegraphics[width=4cm,height=3cm]{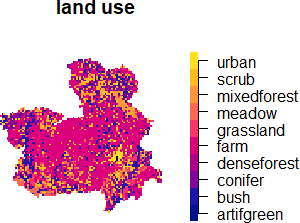}
\includegraphics[width=3.25cm,height=3cm]{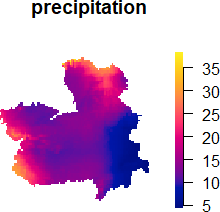}
\includegraphics[width=3.25cm,height=3cm]{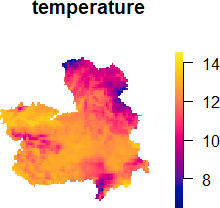}
\caption{\footnotesize{Image plot of environmental covariates (\textit{elevation}, \textit{orientation}, \textit{slope} and \textit{land use}) and climate covariates (\textit{precipitation} and \textit{temperature}) in January $2007$. }}
\label{fig:covclm}
\end{center}
\end{figure}

\noindent
The \texttt{clmfires} dataset available in  \texttt{spatstat} records the occurrences of forest fires in the region of Castilla-La Mancha, Spain (Figure \ref{fig:clm}, left) from 1998 to 2007. 
The study area is approximately $400~km \times 400~km$.
The \texttt{clmfires} dataset has already been used in some academic works devoted to the point process theory (see e.g. Juan et al., 2010; Gomez-Rubio, 2020, sect. 7.4.2; Myllymäki et al., 2020; Kelling and Haran, 2022).
The dataset has two levels of precision: from 1998 to 2003 locations were recorded as the centroids of the corresponding “district units”, while since 2004 locations correspond to the exact UTM coordinates of the fire locations.
Due to the low precision of fire locations for the years 1998 to 2003 (Gomez-Rubio 2020, sect. 7.4.2), we focus on fires in the period 2004 to 2007. 
In this period, we consider large forest fires with burnt areas larger than  $5~ha$.  
Figure \ref{fig:clm} (middle) shows the point pattern of 432 wildfire locations onto the spatial region.  
We consider monthly records; the temporal component of the process lies in $T = \{1,2,\dots,48\}$, where 1 corresponds to January 2004.
Figure \ref{fig:clm} (right) shows a seasonal effect with notably
large numbers of fires in summer that could be caused by high temperatures and low precipitations in this period and also by human activities.
As the spatio-temporal inhomogeneity is notably driven by covariates, 
we include in our analysis four spatial covariates: elevation, orientation, slope and land use (available in  \texttt{spatstat}) and two spatio-temporal covariates:  monthly maximum temperatures ($^{\circ}C$) and total precipitations ($mm$), available on \textit{WorldClim} database\footnote{{\tt www.worldclim.org}}. 
The covariates are known on a spatial grid with pixels of $4~km \times 4~km$, resulting in a total of $10,000$ pixels.
Figure \ref{fig:covclm} illustrates the environmental covariates, which are considered fixed during our temporal period, and the climate covariates in January $2007$.

\medskip

\textbf{\texttt{Prométhée} dataset}

\begin{figure}
\begin{center}
\includegraphics[width=12cm]{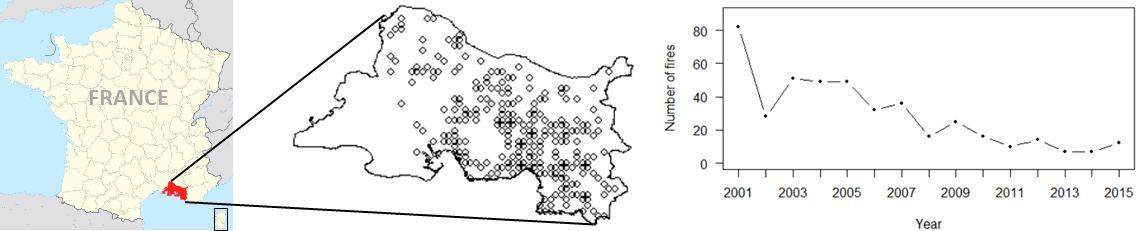}
\caption{\footnotesize{Left: Map of Bouches-du-Rhône (France). Middle: Forest fire locations. Right:  the yearly number of  fires recorded between  $2001$ and  $2015$ with burnt areas, spatial distances, and time distances respectively bigger than 1 $ha$.}}
\label{fig:bdr}
\end{center}
\end{figure}

\noindent
The \texttt{Prométhée} database\footnote{{\tt www.promethee.com}}  was created  to gather several informations relative to the wildfire in French Mediterranean regions. It uses a system of coordinates specially designed for fire management: the DFCI coordinates (DFCI is the French acronym for ”Défense de la Foret Contre les Incendies”, i.e. it refers to all processes concerning forest defense facing to wildfires). 
The fire ignition locations are therefore available with a spatial resolution of 4
$km^2$. The \texttt{Prométhée} has already been used in various academic works devoted to the point process theory, see for example Gabriel et al. (2017); Opitz et al. (2020); Baile et al. (2021); Raeisi et al. (2021); Pimont et al. (2021).
Here, we consider yearly records of forest fire occurrences between 2001 and 2015 in the  Bouches-du-Rhône department (figure~\ref{fig:bdr}, left), with burnt areas larger than 1~$ha$,
and we set $T = \{1,2,\dots,15\}$, where 1 corresponds to 2001.
It contains 434 occurrences (Figure~\ref{fig:bdr}, middle), with a decreasing temporal trend (Figure~\ref{fig:bdr}, right). The locations of fire ignition correspond to the centroid of a grid cell. This induces a hardcore structure in  space and time. As Raeisi et al. (2021), we consider four spatial covariates: water coverage ratio, elevation, coniferous coverage ratio, building coverage ratio, and two spatio-temporal covariates: temperature and precipitation average. We refer the reader to Gabriel et al. (2017), Opitz et al. (2020) and Raeisi et al. (2021)  for well-detailed information on the data and covariates.

\subsection{Inference} 
We consider an inhomogeneous hybrid Strauss hardcore model with Papangelou conditionl intensity 
\begin{equation*}
\lambda((u,v)|\textbf{x}) =\lambda(u,v)\prod_{j=1}^{m} \gamma_j^{n[C_{r_j}^{q_j}(u,v);\textbf{x}]}\prod_{(\xi,t)\in \textbf{x}}\mathbb{1}\{(\xi,t)\notin C_{h_s}^{h_t}(u,v)\},
\end{equation*}
where the spatio-temporal trend is log-linearly related to spatial ($Z_k^S$) and spatio-temporal ($Z_l^{ST}$) covariates as in
\begin{equation} \label{eq:intpois}
    \lambda(\xi,t) = \exp \left( \beta_0 + \sum_{k=1}^4 Z_k^S(\xi) +
    \sum_{l=1}^2 Z_l^{ST}(\xi,t) \right).
\end{equation}
For \texttt{Promethee} we have also a term $\alpha t$. The inference of the hybrid Strauss hardcore model involves different approaches according to the kind of parameters: empirical and ad-hoc for the irregular parameters, likelihood-based for the regular parameters. 

\medskip

\textit{Irregular parameter estimation}

\begin{figure}
\begin{center}
\includegraphics[width=7cm,height=6cm]{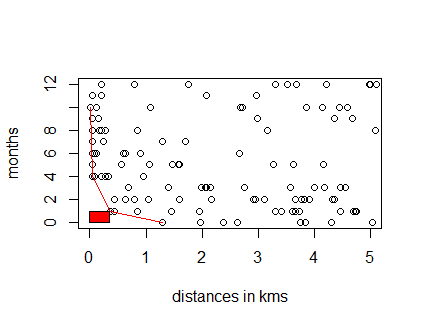}
\caption{\footnotesize{Spatial and temporal interpoint distances respectively lower than $5$ kms and $12$ months (black circles). The red line corresponds to the Pareto front and the red rectangle to the hardcore domain.} }
\label{pic666}
\end{center}
\end{figure}

\noindent
There are two types of irregular parameters: the hardcore distances and the nuisance parameters. 
The hardcore distances can be chosen among all feasible solutions on the Pareto front of spatial and temporal interpoint distances.
For \texttt{clmfires} dataset, according to Figure~\ref{pic666}, we choose on the Pareto front the  feasible solution in our case that gives non-zero values for the two hardcore distances i.e. $h_s=0.35$ $km$ and  $h_t=1$ $month$. For \texttt{Prométhée} dataset, due to the nature of the dataset, we consider $h_s=2000$ $meter$ and $h_t=1$ $year$.  
There is no common method for estimating  the nuisance parameters $m$, $r_j$ and $q_j$, $j=1,\dots, m$. 
A preliminary spatio-temporal exploratory analysis of the interaction ranges based on the inhomogeneous pair correlation function, the maximum nearest neighbor distance and the temporal auto-correlation function, allowed us to set $25$ configurations of reasonable range for the nuisance parameters. We select the optimal irregular parameters according to the Akaike's Information Criterion (AIC) of the fitted model after the regular parameter estimation step (Raeisi et al., 2021). 
 
\medskip

\textit{Regular parameter estimation}

\noindent
 We consider the logistic likelihood method investigated in Section 3 to estimate the regular parameters. 
 The dummy points are simulated from a Poisson process with intensity $\rho(\xi,t)= C \lambda_{\cal P}(\xi,t)/v$, with $C=4$ by a classical rule of thumb in the logistic likelihood approach and $v$ is volume of spatio-temporal region. 
 In order to satisfy the hardcore condition in \eqref{Eq.201322}, we remove dummy points at spatial and temporal distances respectively less than $h_s$ and $h_t$. 

\medskip

\textit{Results}

\noindent
For the \texttt{clmfires} (resp. \texttt{Promethee}) data we select $m=6$ (resp. $m=4$) spatial and temporal interaction distances. Table~\ref{tab:paramclm}  (resp. Table~\ref{tab:paramcbdr}) provides them and the related regular parameters.
\begin{table}
\centering
\caption{Estimated parameters of the hybrid Strauss hardcore model for the \texttt{clmfires} data.}
\label{tab:paramclm}
\scriptsize
\begin{tabular}{rrr}
\hline
 Coefficient & Estimate & $p$-value \\
 \hline
 Intercept  $\beta_0$ &  -24.81  &  $<2\times10^{-16\hspace{2mm}***}$  \\
 Elevation $\beta_1$ &  0.41  &  $0.07846^{\hspace{2mm} .}$ \\
 Orientation  $\beta_2$  & -0.0004   & 0.91913   \\
 Slope  $\beta_3$   &  -0.02 & $0.04898^{\hspace{2mm}*}$   \\
  Land use $\beta_4$  & -0.03   & 0.32856   \\
 Precipitation  $\beta_5$  &  -0.01  &  $2.04\times10^{-7\hspace{2mm}***}$  \\
Temperature  $\beta_6$   &  0.09  &  $<2\times10^{-16\hspace{2mm}***}$  \\
\end{tabular}
\begin{tabular}{rr|rrr}
\hline
r & q & Coefficient & Estimate & $p$-value \\
 \hline
  $r_1=0.5$ &  $q_1=2$   & $\gamma_1$ &  3.04  & 0.49299 \\
 $r_2=1$ & $q_2=4$  &$\gamma_2$ &  3.85  &  $0.04370^{\hspace{2mm}*}$ \\
 $r_3=1.5$ &  $q_3=6$   & $\gamma_3$ & 4.34   & $3.64\times10^{-5\hspace{2mm}***}$  \\
  $r_4=6$ & $q_4=8$  & $\gamma_4$ &  1.01  & 0.94889  \\
  $r_5=15$ &  $q_5=12$   & $\gamma_5$ &  1.19  & $0.00046^{\hspace{2mm}***}$  \\
  $r_6=20$ & $q_6=15$  & $\gamma_6$ &  1.06  & $0.03420^{\hspace{2mm}*}$   \\
\end{tabular}

\end{table}

\begin{table}
\centering
\caption{Estimated parameters of the hybrid Strauss hardcore model for the \texttt{Promethee} data.} 
\label{tab:paramcbdr}
\scriptsize
\begin{tabular}{rrr}
\hline
 Coefficient & Estimate & $p$-value \\
 \hline
Intercept  $\beta_0$  &  79.437  &  0.183666  \\
 Coniferous $\beta^{S}_1$ &  1.342  &  $0.032657^{\hspace{2mm} *}$ \\
 Water $\beta^{S}_2$    & -2.737   & $2.43\times10^{-8\hspace{2mm}***}$ \\
 Building $\beta^{S}_3$  &  -0.885 & 0.647172    \\
 Elevation $\beta^{S}_4$   & -0.003   & $0.000143^{\hspace{2mm}***}$ \\
Temperature $\beta^{ST}_5$  &  0.566  &  $1.33\times10^{-8\hspace{2mm}***}$  \\
Precipitation $\beta^{ST}_6$   &  -28.011  &  $<2\times10^{-16\hspace{2mm}***}$  \\
Time $\alpha$  &  -0.067  &  $0.023502^{\hspace{2mm}*}$ \\
\end{tabular}
\begin{tabular}{rr|rrr}
\hline
r & q & Coefficient & Estimate & $p$-value \\
 \hline
 $r_1=2500$ &  $q_1=2$   & $\gamma_1$ &  0.42  & $0.000141^{\hspace{2mm}***}$  \\
 $r_2=3000$ & $q_2=3$  & $\gamma_2$ &  2.31  &  $1.23\times10^{-8\hspace{2mm}***}$ \\
  $r_3=5000$ &  $q_3=4$   & $\gamma_3$ & 1.26   & $0.020336^{\hspace{2mm}*}$  \\
 $r_4=7000$ & $q_4=5$  & $\gamma_4$ &  0.77  & $2.79\times10^{-6\hspace{2mm}***}$  \\
\end{tabular}

\end{table}

\subsection{Model validation}

\begin{figure}[h]
\begin{center}
\includegraphics[width=.4\textwidth]{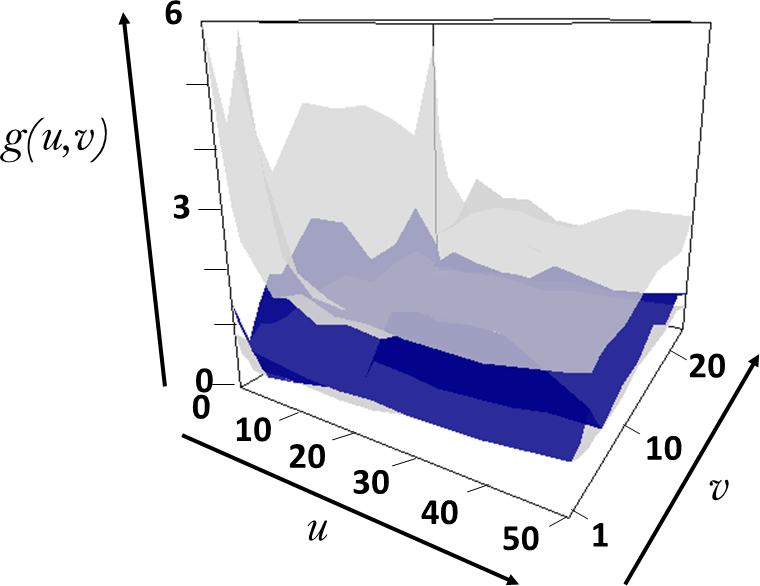}
\caption{\footnotesize{
Envelopes of the spatio-temporal inhomogeneous $g$-function obtained from simulations of the fitted spatio-temporal hybrid Strauss hardcore point process (light grey). The blue surface corresponds to $g_{obs}$. Temporal separations $v$ are in $month$ and spatial distances $u$ are in $kilometer$.}}
\label{pic15}
\end{center}
\end{figure}

The goodness-of-fit of the hybrid Strauss hardcore models is accomplished by simulations from the fitted model. The first diagnostic can be formulated by summary statistics of point processes.  As the second-order characteristics carry most of the information on the spatio-temporal structure (Stoyan, 1992 ; Gonzalez et al., 2016), we only consider the pair correlation function ($g$-function). 
We generate $n_{sim} = 99$ simulations from the fitted hybrid Strauss hardcore models and compute the corresponding second-order summary statistics 
 $g_i(u,v)$, $i =1,\dots,n_{sim}$, for fixed spatio-temporal distances $(u,v)$. We then build upper and lower envelopes:
\begin{equation}
U(u,v)=\max_{1\leq i \leq n_{sim}}  g_i(u,v),  \hspace {.3cm} L(u,v)=\min_{1\leq i \leq n_{sim}}  g_i(u,v), 
\label{Eq.40}
\end{equation}
and compare the summary statistics obtained from the data, $g_{obs}(u,v)$, to the pointwise envelopes. If it lies outside the envelopes at some spatio-temporal distances $(u,v)$, 
 then we reject at these distances the hypothesis that our data come from our fitted model. 
Figure~\ref{pic15} shows the spatio-temporal inhomogeneous $g$-function computed on our dataset from \texttt{clmfires} (blue) and the envelopes obtained from the fitted model (light grey);  $g_{obs}(u,v)$ lies inside the envelopes for all $(u,v)$, meaning that the hybrid Strauss hardcore model is suitable for the data. 

In addition, we compute global envelopes and $p$-value of the spatio-temporal $g$-functions based on the Extreme Rank Length (ERL) measure defined in Myllymäki et al. (2017) and implemented in the \texttt{R} package \texttt{GET} (Myllymäki and Mrkvička, 2020). 
For each point pattern, we consider the long vector $T_i$, $i=1,\dots,n_{sim}$ (resp. $T_{obs}$) merging the $g_i(\cdot,v)$ (resp. $g_{obs}(\cdot,v)$) estimates for all considered values $h_t$. The ERL measure of vector $T_i$ (resp. $T_{obs}$) of length $n_{st}$ is defined as
$$ E_i = \frac{1}{n_{ns}} \sum_{j=1}^{n_{st}} \mathbbm{1} \lbrace R_j \prec R_i \rbrace,$$
where $R_i$ is the vector of pointwise ordered ranks and $\prec$ is an ordering operator (Myllymäki et al., 2017; Myllymäki and Mrkvička, 2020). The final  $p$-value is obtained by
$$p_{erl}=\frac{1 + \sum_{i=1}^{n_{sim}} \mathbbm{1} \lbrace E_i \geq E_{obs} \rbrace}{n_{sim}+1}.$$

\begin{figure}[h]
\begin{center}
\includegraphics[width=1\textwidth]{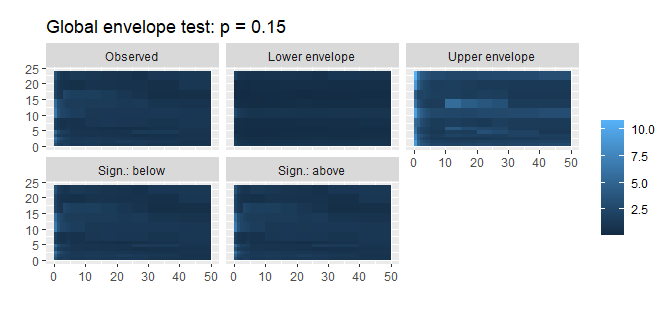}
\caption{\footnotesize{Top: estimated pair correlation function $\hat{g}_{obs}$, lower $E_L$ and upper $E_U$ bounds of the $99 \%$ global rank envelope (ERL). Bottom: differences $E_{obs} - E_L$ and $E_U - E_{obs}$  related to  \texttt{clmfires}. Negative values (if any) are represented in red and lead to reject the fitted model. Values on the horizontal axis are in kilometers and those on the vertical axis are in months.}}
\label{pic160}
\end{center}
\end{figure}

\begin{figure}[h]
\begin{center}
\includegraphics[width=1\textwidth]{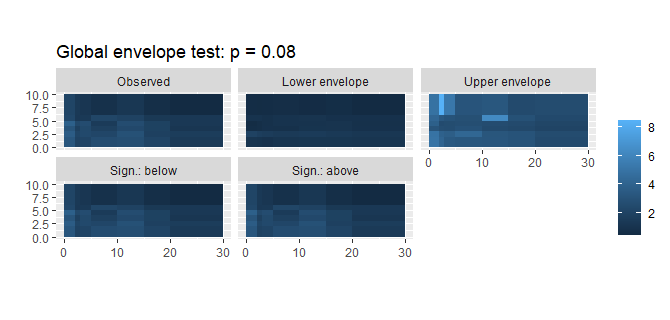}
\caption{\footnotesize{Top: estimated pair correlation function $\hat{g}_{obs}$, lower $E_L$ and upper $E_U$ bounds of the $99 \%$ global rank envelope (ERL). Bottom: differences $E_{obs} - E_L$ and $E_U - E_{obs}$ related to \texttt{Prométhée}. Negative values (if any) are represented in red and lead to reject the fitted model. Values on the horizontal axis are in kilometers and those on the vertical axis are in years.}}
\label{pic16}
\end{center}
\end{figure}

For \texttt{clmfires} and \texttt{Prométhée}, due  to  the global $p$-values $p_{erl}=0.15,0.08$ respectively  and the absence of significant  regions,
 that corresponds here to pairs of spatial and temporal distances where the statistics is significantly above or below the envelopes (see Figure~\ref{pic160}, Figure~\ref{pic16}  and \texttt{GET} package), we conclude that our  hybrid  Strauss  hardcore  model can not be rejected.

\subsection{Model comparison}
We are interested to evaluate the performance of the hybrid Strauss hardcore model over other models. Hence, we fit an inhomogeneous Poisson point process and a LGCP to two wildfire datasets.

\subsubsection{Poisson  process} 
\begin{figure}[h]
\begin{center}
\includegraphics[width=1\textwidth]{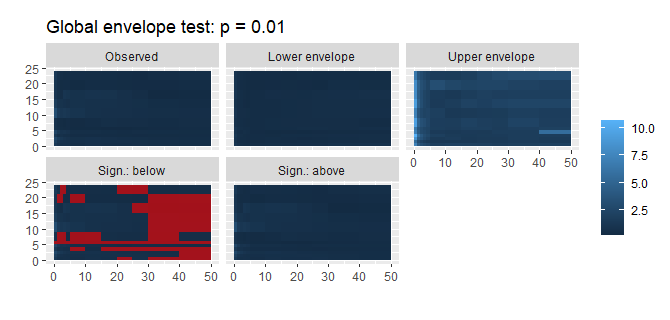}
\includegraphics[width=1\textwidth]{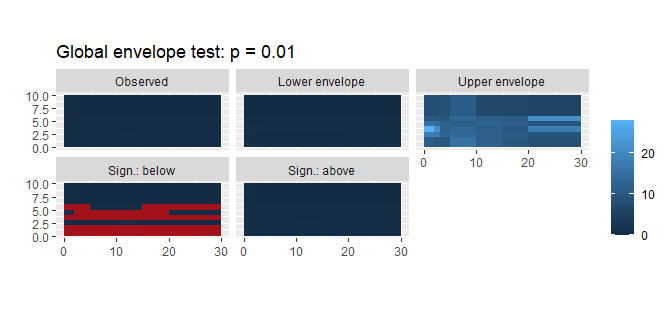}
\caption{\footnotesize{Estimated pair correlation function $\hat{g}_{obs}$, lower $E_L$ and upper $E_U$ bounds, differences $E_{obs} - E_L$ and $E_U - E_{obs}$ of the $99 \%$ global rank envelope (ERL) from two inhomogeneous Poisson point processes related to  \texttt{clmfires} (up) and \texttt{Prométhée} datasets (down).}}
\label{pic150}
\end{center}
\end{figure}

To evaluate the performance of the hybrid Strauss hardcore model, we generated  99 simulations from two inhomogeneous spatio-temporal Poisson point processes with estimated intensity $\lambda/(4\times 4 \times 1)$  for \texttt{clmfires} dataset and $\lambda/(2000\times 2000 \times 1)$  for \texttt{Prométhée} dataset by \texttt{rpp} function in \texttt{stpp} package (Gabriel et al., 2013). Figure \ref{pic150} is the plots of upper and lower of the global $p$-value for  99 simulated Poisson point process patterns with value 0.01 related to \texttt{clmfires} (up) and \texttt{Prométhée} datasets (down)  that confirms the advantage the hybrid Strauss hardcore point process over the inhomogeneous Poisson point process for our two forest fire datasets.

\subsubsection{LGCP} 
For more comparison, we fit a LGCP on  \texttt{clmfires} and \texttt{Prométhée} datasets by INLA approach. We consider a LGCP with stochastic intensity consists a space-month resolution for incorporating covariate information. The model is considered for a Gaussian space-time effect $W(\xi, t)$, whose spatial component is
always based on the flexible yet computationally convenient Matérn-like spatial Gauss–Markov
random fields arising as approximate solutions to certain stochastic partial differential equations. By adding the Gaussian space-time effect $W(\xi, t)$, we define a LGCP with a stochastic intensity $\log \Lambda(\xi,t)$ rather than $\log \lambda(\xi,t)$. 
\begin{equation*} 
    \log \Lambda (\xi,t) = \beta_0 + \sum_{k=1}^4 Z_k^S(\xi) +
    \sum_{l=1}^2 Z_l^{ST}(\xi,t) + W(\xi,t),
\end{equation*}
For \texttt{Promethee} we have a term $\alpha t$. The inference is based on the Integrated Nested Laplace Approximation (Rue et al., 2009) and its implementation in \texttt{R-INLA}. We refer the reader to Opitz \textit{et al.}~(2020) for a thorough description of the implementation of very similar models and data. 

\begin{figure}
	\begin{center}
		\includegraphics[width=1\textwidth]{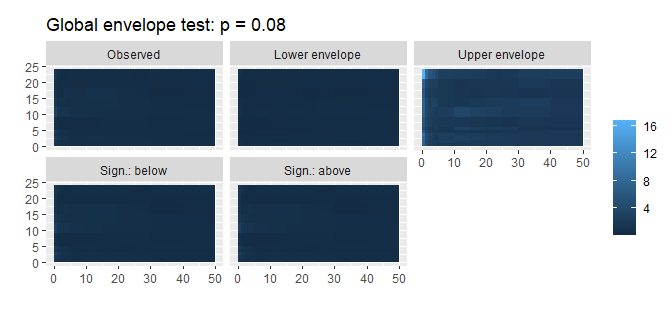}
		\includegraphics[width=1\textwidth]{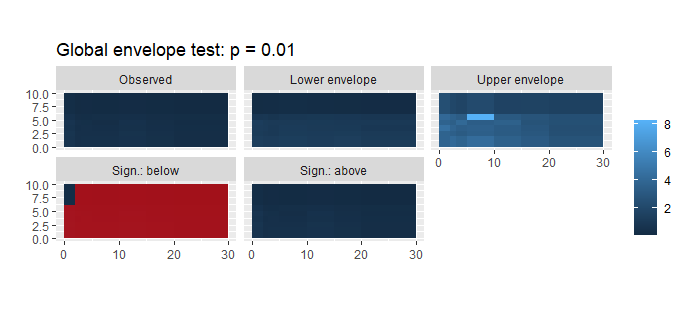}
		\caption{Estimated pair correlation function $\hat{g}_{obs}$, lower $E_L$ and upper $E_U$ bounds, differences $E_{obs} - E_L$ and $E_U - E_{obs}$ of the $99 \%$ global rank envelope (ERL) from two LGCPs related to  \texttt{clmfires} (up) and \texttt{Prométhée} datasets (down).}
		\label{pic20201}
	\end{center}
\end{figure}

Figure \ref{pic20201} is the plots of upper and lower of the global $p$-value for  99 simulated point patterns from LGCP with values $0.08,0.01$ related to \texttt{clmfires} (up) and \texttt{Prométhée} datasets (down)  that confirms the advantage the hybrid Strauss hardcore point process over the LGCP for our two forest fire datasets.

In summary, the global envelope tests may be used for model comparison by comparing $p$-values and concluding that the model with the highest $p$-value provides the best fit.
Use consider this test to compare the Inhomogeneous Poisson, Log-Gaussian Cox, and hybrid Strauss hardcore processes on the two datasets. The $p$-values are given in Table~\ref{tab:modcomp} that is shown  a good candidate can be Gibbs point processes based on hybridization of Strauss and hardcore densities to model the forest fire occurrences.
\begin{table}[h]
    \centering
    \begin{tabular}{r|ccc}
         Model & IPP & LGCP & SH \\
         \hline
         \texttt{clmfires} & 0.01 & 0.08 & 0.15 \\
         \texttt{Promethee} & 0.01 & 0.01 & 0.08 \\
    \end{tabular}
    \caption{Model comparison based on $p_{erl}$.}
    \label{tab:modcomp}
\end{table}

\section*{Conclusion}
In this paper, we introduced the spatio-temporal Strauss hardcore point process. The Strauss hardcore model is a Gibbs model for which points are pushed to be at a hardcore distance apart and repel up to a interaction distance  which is larger than the hardcore distance. As in Raeisi et al. (2021), inference and simulation of the model were performed with logistic likelihood and birth-death Metropolis-Hasting algorithm, respectively. A multi-scale version of the model was introduced and applied to wildfires to take into account structural complexity of forest fire occurrences in space and time. We based our model validation on both pointwise and global envelopes and $p$-value based on the Extreme Rank Length (ERL) measure of the spatio-temporal inhomogeneous pair correlation function.

Note that, here, our criterion for choosing the best fitted model was to compare the extracted AIC value of the GLM of conditional intensity implemented in \texttt{R}. The composite AIC criterion for spatial Gibbs models introduced by Daniel et al. (2018) and Choiruddin et al. (2021) could have been better. However, the composite AIC requires to estimate the variance–covariance matrix of the logistic score and the sensitivity matrix which should be extended to the spatio-temporal framework  that is a full-blown work and also involves efficient implementation.

Our model could be suitable in other environmental and ecological frameworks, when we want to deal with the complexity of mechanisms governing attraction and repulsion of entities (particles, cells, plants$\dots$).

In spatio-temporal  Gibbs point process models, the heterogeneity  can be captured by estimating a non-constant  trend. This spatio-temporal trend is typically considered as a function of covariates  by estimating   fixed effects in a generalized linear model as we carried out it in this paper and also done in Iftimi et al. (2018) and Raeisi et al. (2021).
A different approach consists in considering Gibbs models with both random and fixed effects (e.g. see Illian and Hendrichsen, 2010) to take into account complex patterns of spatio-temporal heterogeneity. Vihrs et al. (2021)  proposed a new modeling approach for this case and embedded spatially structured Gaussian random effects in trend function of a pairwise interaction process. They introduced the spatial log-Gaussian Cox Strauss point process to capture both structures; aggregation in small-scale and repulsion in large-scale.  Rather than spatial pairwise interaction processes in single-scale, we can focus on  models derived from the multi-scale classes of combinations of Gibbs and log-Gaussian Cox point processes in space and time. Indeed, in a work in progress, we aim to propose to embed spatio-temporally structured Gaussian random effects in the Gibbs trend function.  Due to the hierarchical structure of such models, we can formulate and estimate them within a  Bayesian hierarchical framework, using the INLA approach.

\end {document}